\documentclass[final,3p,times,twocolumn]{article}

\usepackage{PRIMEarxiv}

\usepackage[utf8]{inputenc} 
\usepackage[T1]{fontenc}    
\usepackage{hyperref}       
\usepackage{url}            
\usepackage{booktabs}       
\usepackage{amsfonts}       
\usepackage{nicefrac}       
\usepackage{microtype}      
\usepackage{lipsum}
\usepackage{fancyhdr}       
\usepackage{graphicx}       
\graphicspath{{media/}}     

\usepackage{amssymb}
\usepackage{amsmath}
\usepackage[flushleft]{threeparttable}
\usepackage{booktabs,caption}
\usepackage[labelfont=bf]{caption}
\usepackage{subcaption}
\usepackage{multirow}
\usepackage[version=4]{mhchem}
\usepackage{parskip}
\usepackage{authblk}
\usepackage[toc]{appendix}
\usepackage{xpatch}
\usepackage{cite}

\title{NMR spectrum reconstruction as a pattern recognition problem
\thanks{\textit{A Preprint}}
}

\author[a$\dagger$]{\textbf{Amir Jahangiri}}
\author[b]{\textbf{Xiao Han}}
\author[c]{\textbf{Dmitry Lesovoy}}
\author[d]{\textbf{Tatiana Agback}}
\author[d]{\textbf{Peter Agback}}
\author[b]{\textbf{Adnane Achour}}
\author[a$\dagger$]{\textbf{Vladislav Orekhov}}

\affil[a]{Department of Chemistry and Molecular Biology, Swedish NMR Centre, University of Gothenburg, Box 465, Gothenburg, 40530, Sweden
\thanks{\textit{Corresponding authors: \href{mailto:amir.jahangiri@gu.se}{amir.jahangiri@gu.se} (Amir Jahangiri),
\href{mailto:vladislav.orekhov@nmr.gu.se}{vladislav.orekhov@nmr.gu.se}}(Vladislav Orekhov)}}
\affil[b]{Science for Life Laboratory, Department of Medicine, Karolinska Institute, and Division of Infectious Diseases, Karolinska University Hospital, Stockholm, 17176, Sweden}
\affil[c]{Shemyakin-Ovchinnikov Institute of Bioorganic Chemistry RA, Moscow, 117997, Russia}
\affil[d]{Department of Molecular Sciences, Swedish University of Agricultural Sciences, Box 7015, Uppsala, 75007, Sweden}

\begin{document}

\onecolumn
\maketitle

\begin{center}
\today  
\end{center}

~

\begin{abstract}
A new deep neural network based on the WaveNet architecture (WNN) is presented, which is designed to grasp specific patterns in the NMR spectra. When trained at a fixed non-uniform sampling (NUS) schedule, the WNN benefits from pattern recognition of the corresponding point spread function (PSF) pattern produced by each spectral peak resulting in the highest quality and robust reconstruction of the NUS spectra as demonstrated in simulations and exemplified in this work on 2D \ce{^{1}H}-\ce{^{15}N} correlation spectra of three representative globular proteins with different sizes: Ubiquitin (8.6 kDa), Azurin (14 kDa), and Malt1 (44 kDa). The pattern recognition by WNN is also demonstrated for successful virtual homo-decoupling in a 2D methyl \ce{^{1}H}-\ce{^{13}C} \textemdash\ HMQC spectrum of MALT1. We demonstrate using WNN that prior knowledge about the NUS schedule, which so far was not fully exploited, can be used for designing new powerful NMR processing techniques that surpass the existing algorithmic methods.
\end{abstract}


\twocolumn
\section{Introduction}
\label{sec:Introduction}

    NMR spectroscopy is an analytical technique that provides atomic-level information about molecular structure, dynamic, and interactions \cite{claridge2016high,cavanagh1996protein}. Since the invention of Fourier NMR in 1970's, modern spectra are acquired by sampling the signal in the domain. This not only significantly improved sensitivity, but also enabled rapid development of multidimensional NMR experiments that offer ultimate resolution and rich information content. However, increase in resolution and dimensionality leads to very long measurement time needed to systematically Uniformly Sample (US) large volume of the multidimensional data set \cite{Jaravine_2006}. To address this problem, the major fraction of the US data is not measured in the approach called Non-Uniform Sampling (NUS), mainly used in modern experiments. Since Fourier transform can no longer produce high quality spectra reconstruction from the NUS data, many alternative signal processing techniques were developed in the past years \cite{Jaravine_2006,MOBLI2014,Qu2015,Kazimierczuk2011,Hyberts2012,Hassanieh2015,Hyberts2010,Pustovalova2018,sibisi1984maximum,drori2007fast,holland2011fast,JIANG2010165,Ying2017SparseMI}. Reconstruction of a spectrum from NUS, \textit{i.e.} incomplete data, is an ill-defined mathematical problem which can only be solved by introducing prior assumptions about the spectrum. For example, in the approach known as Compressed Sensing, the most sparse spectrum is selected \cite{Kazimierczuk2011}. Generally, the more we constrain the solution with correct assumptions, the better is the reconstruction quality and the less NUS data are needed. Until recently, the progress in the NUS spectra processing focused on solving two practical tasks, specifically to \textit{i)} define the best priors and \textit{ii)} design computationally effective algorithms for their implementation. 
    
    Artificial Intelligence (AI) and specifically Deep learning (DL) has a potential to solve both the above-mentioned tasks in a new and efficient way. Although the first demonstrations of the machine learning applications in NMR can be traced back to 1970s \cite{reilly1971nuclear}, practical applications could not be developed until reaching the modern level of algorithms and computer hardware. Over the last years, DL has led to impressive advances in many fields, including NMR spectroscopy \cite{Chen2020}. For example, DL has a marked ability for the spectra denoising \cite{Lee2019}, prediction of chemical shift \cite{Liu2019}, and performing automated peak picking \cite{Klukowski2018, li2022fundamental}, as well as for fast and high quality NMR reconstruction of NUS spectra \cite{Qu2019,Hansen2019,Karunanithy2021,LUO2020106772}. The distinctive feature of DL neural networks (DNNs) is their ability to establish essential correlations between the input and output and thus to retrieve relevant multi-facet priors, \textit{e.g.} about NMR signal, that are difficult to formulate in an analytical form and embed into a computer algorithm. Although a trained DNN is usually considered as a black box, there are reports, where the knowledge mined by the network in the training process was rationalized and even reverse-engineered \cite{amey2021neural}. In this work, we learned from the DNN that the NUS schedule is also a valuable and so far not fully exploited source of prior information for the spectrum reconstruction. DNNs are generally very efficient in pattern recognition \cite{Pattern}. In NMR, DNNs were successfully used for automated peak picking \cite{li2022fundamental} and virtual homo-nuclear decoupling \cite{karunanithy2021virtual}. In this work, we note that very distinct pattern of spectral aliasing artefacts corresponding to a particular sampling schedule can be effectively recognized and rectified by DNN. In this study, we develop and train new deep neural networks to solve the NUS reconstruction problem. We call our software tool – WNN as it is inspired by WaveNet DNN architecture which was originally conceived in 2016 as a model for raw audio signal \cite{Wave_net}. The broad reception field featured by the WaveNet DNNs is particularly important for grasping the entire pattern of NUS-associated aliasing artefacts, which spread over the entire spectrum area. 
    
    Here, after description of the network architecture and synthetic data used for the training, we compared the performance of the different reconstruction protocols for representative two-dimensional NUS spectra of three globular proteins of different sizes. The four reconstruction protocols include Compressed Sensing Iterative Soft Thresholding algorithm (CS-IST) \cite{Kazimierczuk2011} as a representative traditional method, as well as three WNNs differing in data sets used for their training. Specifically, WNN-F trained on a fixed, \textit{i.e.} the same, NUS schedule for all training spectra, WNN-P and WNN-R trained on different NUS tables from Poisson-gap and flat-random distribution, respectively. Finely, as another demonstration of the pattern recognition by WNN, we present the virtual decoupling on the example of a methyl \ce{^{1}H}-\ce{^{13}C} \textemdash\ HMQC spectrum in the 44 kDa protein MALT1 \cite{unnerstaale2016backbone,malt12022met}. 
    
\section{Materials and Methods}
\label{sec:Materials and methods}

    \subsection{Synthetic data for Training and Testing}
    
        \subsubsection{Data Model}
            Obtaining of a very large data set for the DNN training is a challenge. It is common to use easily generated synthetic data as a good proxy for the realistic experimental NMR spectra. In many practical applications NMR time domain signal $X_{FID}$, usually called free induction decay (FID), can be presented as a superposition of a small number of exponential functions:
        
            \begin{equation}
            \small
                X_{FID}(t_1) = \sum_{n} a_n e^{-t_1/\tau_n} e^{i(2 \pi \omega_n t_1+\phi_n)} cos(\pi J_n t_1) + noise
                \label{eq:exp_fun}
            \end{equation}

            where $n$ runs over the number of exponentials, and the $n$th exponential has the amplitude $a_n$, phase $\phi_n$, relaxation time $\tau_n$, frequency $\omega_n$. In addition, we added in the present work the $J$-coupling constant $J_n$. The time $t_1$ is given by multiplication of the dwell time ($DW$) and the series 0, 1, ..., N-1, which enumerate the sampled time points in the FID. The total acquisition time (AT) is given by product $(N{\times}DW)$. Also, we added the Gaussian $noise$ to emulate the noise presented in realistic NMR spectra. The desired number of different FIDs for the training set is easily simulated by varying the above parameters. The parameters used for generating the synthetic data are summarized in Table. \ref{table:pra_1D}.
            
            \begin{table}[htbp]
                \centering
                \small
                \caption{Parameters for the synthetic FID}
                \begin{tabular}{l| c c}
                    \hline
                    \hline
                     & NUS reconstruction & Virtual decoupling\\
                    \hline
                    $a_n\in \mathbb{R}$ & 0.05 - 1 & 0.05 - 1 \\
                    $\omega_n\in \mathbb{R}$ & 0.1 - 0.9 & 0.1 - 0.9\\
                    $\tau_n\in \mathbb{R}$ & $0.5AT$ - $5AT$ & $0.25AT$ - $0.5AT$\\
                    $\phi_n\in \mathbb{R}$ & $-5^\circ$ - $5^\circ$ & $-5^\circ$ - $5^\circ$ \\
                    $J_n\in \mathbb{R}$ & 0 & $27Hz$ - $45Hz$ \\
                    $N\in \mathbb{N}$ & 128 & 256 \\
                    $n\in \mathbb{Z}$ & 0 - 30 & 0 - 15 \\
                    \hline
                    $noise\in \mathbb{C}$ & 0 - 0.01$^*$ & 0 - 0.01$^*$ \\
                    \hline
                    \hline
                    \multicolumn{3}{l}{\footnotesize $^*$ The absolute value of a complex number} \\
                \end{tabular}
                \label{table:pra_1D}
            \end{table}
    
            The uniformly sampling spectrum is obtained by:
        
            \begin{equation}
                S_{FID}=DFT(X_{FID})
            \label{eq:DFT}
            \end{equation}
        
            where $DFT()$ is the discrete Fourier transform.
        
            Then, the Non Uniform Sampling (NUS) signal, $Y$, is generated from $X_{FID}$ as:
        
            \begin{equation}
                Y=u \circ X_{FID}
            \label{eq:Y}
            \end{equation}
        
            where $\circ$ is Hadamard product and $u$ is the NUS schedule in a vector the representation with ones and zeros at the acquired and non-acquired positions in $X_{FID}$, respectively. 
        
            Using the Fourier convolution theorem, the NUS spectrum, $s_{NUS} = DFT(Y)$ is:
        
            \begin{equation}
                \begin{split}
                    s_{NUS} & = DFT(u \circ X_{FID}) \\
                     & =DFT(u)*DFT(X_{FID})\\
                     &=U*S_{FID}
                \end{split}
            \label{eq:CON}
            \end{equation}
        
            where $U=DFT(u)$ is the Point Spread Function (PSF) and $*$ is convolution. Because of the convolution of $U$ and $S_{FID}$, each signal in the spectrum generates a unique pattern of the random-noise-like artifacts, which is defined by the specific NUS schedule. Note that since function $u$ is real, PSF is symmetric around the true peak position.
            
            ~
            
            In the case of 2D spectra used in this work, the NUS and above-described model equations (\ref{eq:exp_fun}-\ref{eq:CON}) are applied only to the indirectly detected spectral dimension referred as the first with time variable $t_1$. The second, direct dimension, is acquired in full and Fourier transformed. Thus, slices corresponding to the adjacent $\omega_2$ points may share the same signals in $t_1$ dimension. To take this into account, we use a 3-point sliding window in $\omega_2$ dimension. For the points in the window, Eq. \eqref{eq:exp_fun} applies with the same parameters except for the amplitude $a_n$. Since we assume the peaks in the direct dimension as Lorentzian lines, the amplitude $a_n^k$ of exponent $n$ in slice $k$ is:
            
            \begin{equation}
                a^k_n=\frac{\lambda^2}{\lambda^2 + (k \Delta \omega_2)^2}
            \end{equation}
            
            where $\Delta \omega_2$ is the spectrum digital resolution for the directly detected dimension $\omega_2$. $\lambda$ is the line width, which is randomly selected in the range $\Delta \omega_2 < \lambda <5\Delta \omega_2$. Then, $k$ runs over three consecutive integer numbers resulting in $0.05 \leq a^k_n \leq 1$.

        \subsubsection{Training WNN with different NUS-PSF strategies}
    
\begin{figure}[htbp]
                \centering
                \includegraphics[width=0.45\textwidth]{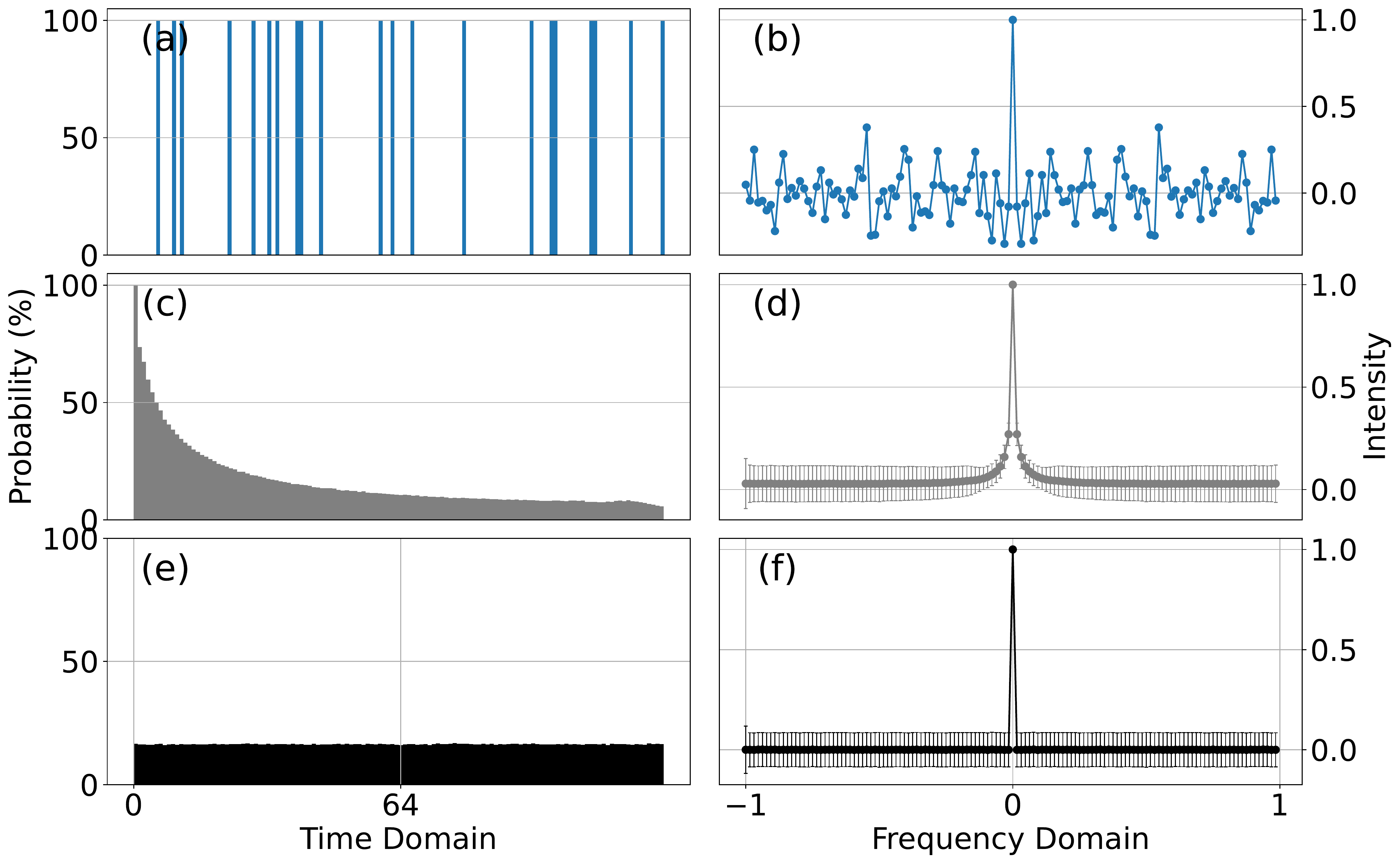}
                \caption{The sampling scheme statistics. (a, c, e) - averaged percentage values for NUS schemes (21 points from the grid of 128) for $2^{16}$ different seeds. (b, d, f) - mean PSF values with standard deviations as the error bars. (a,b) fixed and (c, d) unfixed Poisson-gap NUS/PSF with $sin(\frac{\pi}{2} \frac{t}{t_{max}})$ modulation; and (e, f) NUS/PSF for unfixed random  sampling}
                \label{fig:random}
            \end{figure}

            When generating data sets for training the WNNs, we used three strategies: (i) Fixed (WNN-F), where the same NUS schedule of the Poisson-gap type \cite{Hyberts2010} was used for all $K$ spectra in the training set, (ii) unfixed Poisson-gap (WNN-P), where each spectrum had individual Poisson-gap schedules, and (iii) unfixed random (WNN-R) with the individual NUS tables followed the flat random sampling distribution. Averaged NUS tables and PSFs for the three approaches are presented in Fig. \ref{fig:random}. Implication of the different sampling schemes for appearance of the PSF and for the outcome of the spectra reconstruction using traditional techniques have been previously discussed previously \cite{Kasprzak2021}. Specifically, the Poisson-gap schedules tend to push the aliasing artefacts away from the main peak and thus benefit spectra with clustered signals. When it comes to the WNN, the three approaches clearly differ in their PSF patterns, which can affect pattern recognition by the network. While WNN-F (Fig. \ref{fig:random}b) has a distinct recognizable pattern, which is repeated for each true peak in the spectrum, WNN-R (Fig. \ref{fig:random}f) has the most featureless average PSF, which is the least informative for the pattern recognition. WNN-P (Fig. \ref{fig:random}d) with the visible featured slopes at the base of the central PSF peak takes position in-between WNN-F and WNN-R.
            
            Tables \ref{table:pra_1D} and \ref{table:pra_training} present details of the training set parameters. In this work, we demonstrate WNN performance for several spectra of different complexity representing small, medium, and large proteins. For each of three used proteins, we trained eleven WNN-F with eleven different Poisson-gap sampling schedules, one WNN-P, and one WNN-R. This gives in total 39 WNNs.

            We also trained a WNN for the virtual decoupling (VD) in a methyl 2D \ce{^{1}H}-\ce{^{13}C} \textemdash\ HMQC spectrum, where without the decoupling the \ce{^{13}C} resonances are split due to approximately 35 Hz J-coupling with the adjacent to the methyl \ce{^{13}C} atom. For the VD, the WNN, which contained only one DNN (Fig. \ref{fig:DNN}) and no correction step, was trained on uniform sampling data (see Section \ref{sec:WaveNet-based NMR Network (WNN) Architecture} for details). 
            
            \begin{table}[htbp]
            \centering
            \small
                \caption{Training set parameters}
                \begin{tabular}{l|ccc|c}
                    \hline
                    \hline
                     & \multicolumn{3}{c|}{\textbf{NUS Reconstruction}} & \textbf{VD} \\
                    & Small & Medium & Large & \\
                     
                     \hline
                     \hline
                    $K_{Training}^a$ & $2^{16}$ & $2^{16}$ & $2^{16}$ & $2^{16}$ \\
                    
                    $n_{max}^b$ & 10 & 20 & 30 & $15 \times 2$ \\
                    NUS rate & $\frac{11}{128}$ & $\frac{21}{128}$ & $\frac{31}{128}$ & - \\
                    
                    \# of WNN-F & 11 & 11 & 11 & -\\
                    \# of WNN-P & 1 & 1 & 1 & -\\
                    \# of WNN-R & 1 & 1 & 1 & -\\
                    \hline
                    \textbf{\# of WNNs} & \textbf{13} & \textbf{13} & \textbf{13} & \textbf{1} \\
                    \hline
                    \hline
                    \multicolumn{5}{l}{\footnotesize $^a$ The number of spectra in the training set} \\
                    \multicolumn{5}{l}{\footnotesize $^b$ The maximal number of peaks along indirect dimension} \\
                \end{tabular}
                \label{table:pra_training}
            \end{table}

        \subsubsection{Test data}
        \label{sec:Test data}
            In order to test trained WNNs, we generated synthetic data representing spectra for small, medium, and large proteins. For each case, the test data set contains 1500 uniform sampling synthetic 3-point sliding window spectra (Table \ref{table:pra_testing}): 1 ($\times$ 150) to 10 ($\times$ 150) peaks for a small protein, 1 ($\times$ 75) to 20 ($\times$ 75) peaks for a medium protein, and 1 ($\times$ 50) to 30 ($\times$ 50) peaks for a large protein. By using 11 NUS sampling schemes used for training 11 WNN-F for each protein size, 11 different NUS spectra for each protein size are simulated and then reconstructed. 
            \begin{table}[htbp]
            \centering
            \small
                \caption{Testing set parameters}
                \begin{tabular}{l|ccc}
                    \hline
                    \hline
                    & Small & Medium & Large\\
                     
                     \hline
                     \hline
                    $K_{Testing}^a$ & 1500 & 1500 & 1500\\
                    
                    $n_{max}$ & 10 & 20 & 30\\
                    NUS rate & $\frac{11}{128}$ & $\frac{21}{128}$ & $\frac{31}{128}$\\
                    
                    \hline
                    \hline
                    \multicolumn{4}{l}{\small $^a$ The number of spectra in the testing set} \\

                \end{tabular}
                \label{table:pra_testing}
            \end{table}
            
    \subsection{WaveNet-based NMR Network (WNN) Architecture}
    \label{sec:WaveNet-based NMR Network (WNN) Architecture}
\begin{figure}[htbp]
                \centering
                \begin{subfigure}[b]{0.45\textwidth}
                    \centering
                    \caption{}
                    \includegraphics[width=\textwidth]{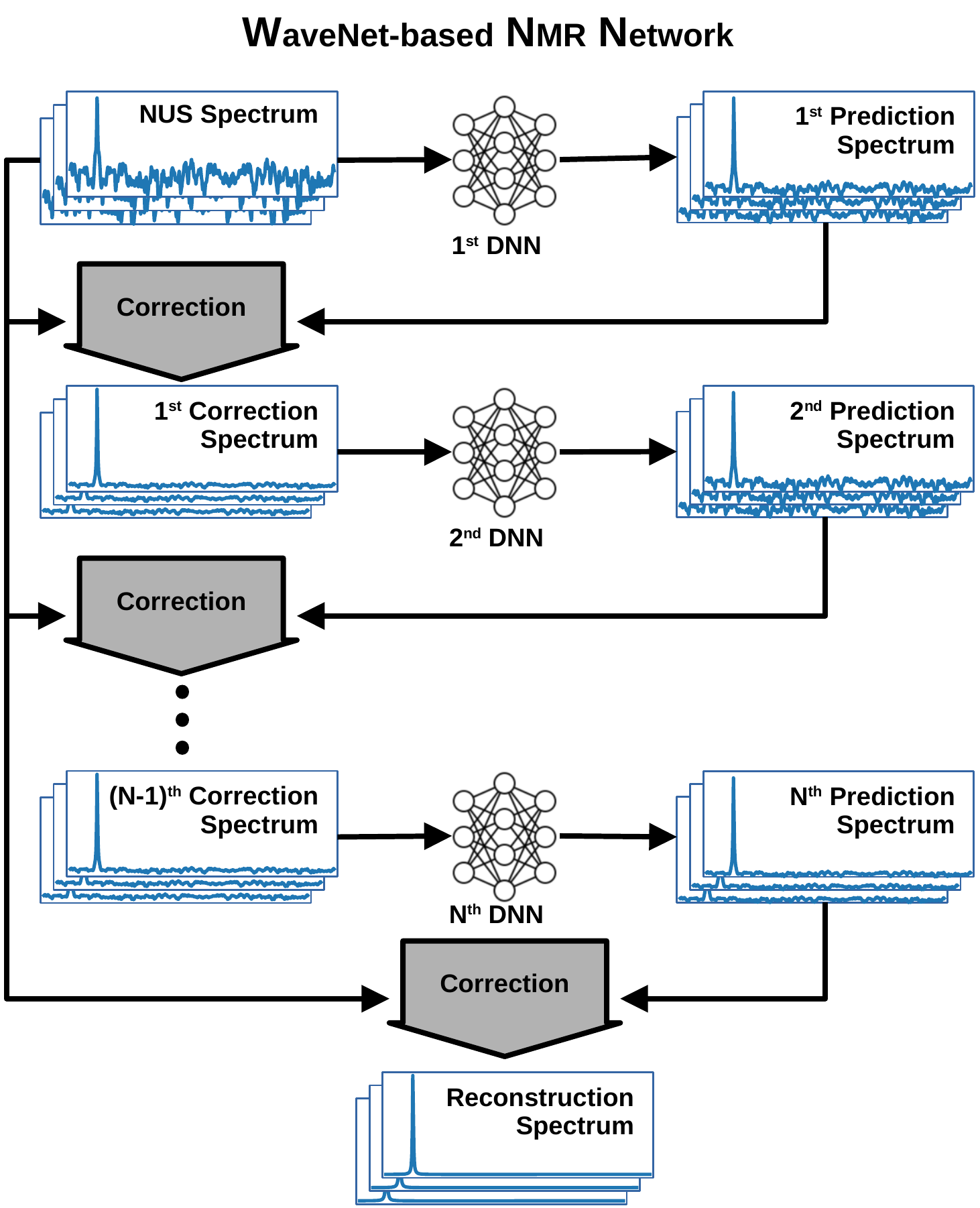}
                    \label{fig:WNN}
                \end{subfigure}
                \begin{subfigure}[b]{0.45\textwidth}
                    \centering
                    \caption{}
                    \includegraphics[width=\textwidth]{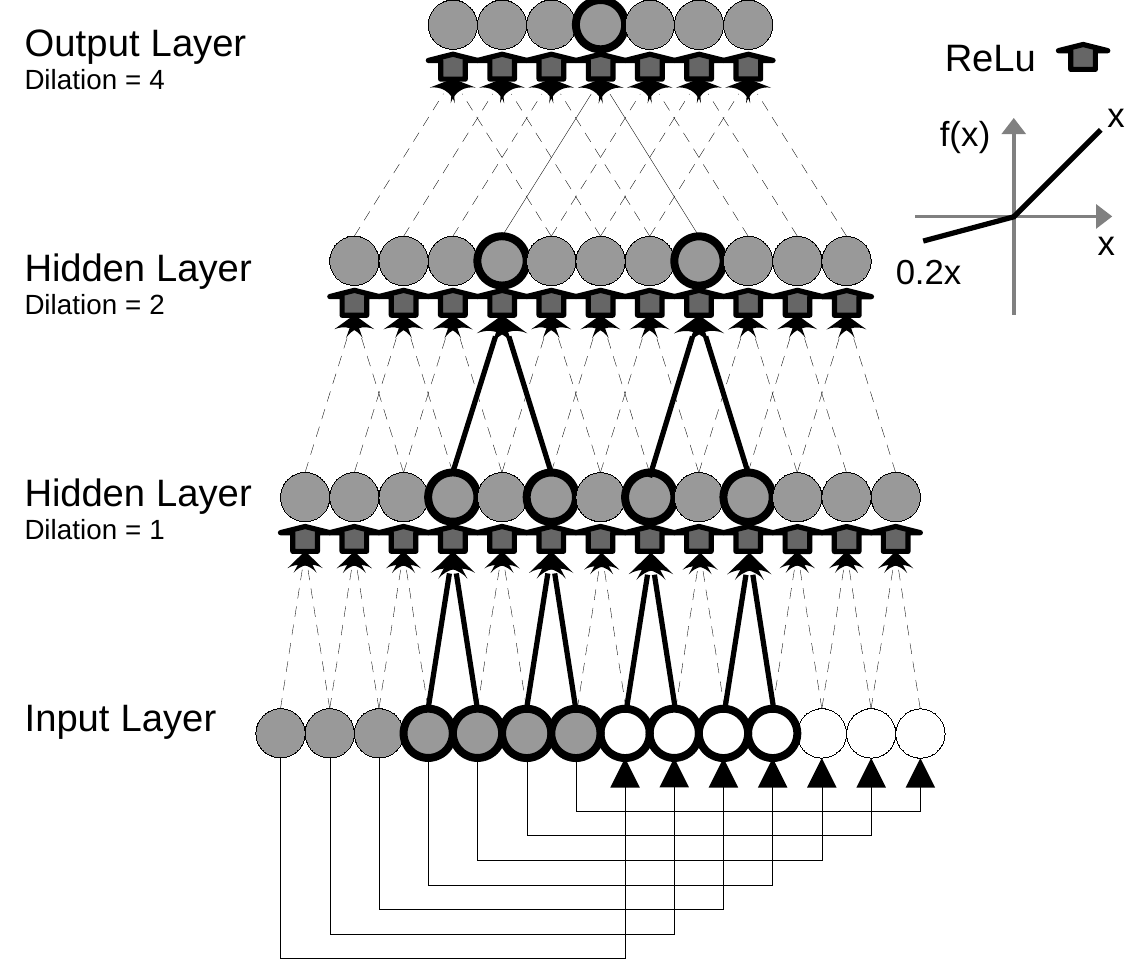}
                    \label{fig:DNN}
                \end{subfigure}
                \caption{ a) WNN network architecture. b) Scheme of the DNN module used in WNN with $2^n-1$ ($n \in \mathbb{N}$) points in the input layer, $n-1$ hidden layers, 20 filters, and ReLU activation function between the layers.}
            \label{fig:WDNN}
            \end{figure}

        In WNN, the entire network architecture consists of three main components: the DNNs, normalization, and the correction steps (Fig. \ref{fig:WDNN}).
        
        \subsubsection{A Specific DNN Architecture Used in WNN}
            Schematics of the WNN are presented in Fig. \ref{fig:WDNN}. For the extraction of information about the PSF patterns within a NUS spectrum (or the J-coupling patterns), a large receptive field is needed. Our DNN is based on the waveNet architecture, which was originally developed for analysis of raw audio signals \cite{Wave_net}. With WaveNet's dilated convolution layers, the network effectively observes the audio signal in a wide receptive field. Similarly, WNN perceives the whole NUS spectrum and can detect the entire PSF patterns. The dilated convolutional layers skip a defined number of points in the data and thus can be considered as convolutional layers with gaps. With different dilation sizes for different convolutional layers, it is possible to build a block that behaves like a convolutional layer with a very large filter size. WNN architecture inherits the idea of the dilated convolutional layers from the WaveNet. There are also a number of significant differences between the WNN and WaveNet architectures. Specifically, while the WNN employs the Rectified Linear Unit (ReLU) activation functions \cite{relu} (Fig. \ref{fig:DNN}), the WaveNet utilizes the gated activation unit, the same as used in the gated PixelCNN \cite{van2016conditional}. WNN does not have the residual or parameterized skip connections \cite{he2016deep}, which are present in WaveNet. Furthermore, WaveNet is designed to be a forward audio generative model, maintaining the temporal order of the data and predicting outputs based only on the preceding values. This requires the use of causal padding in convolutional layers. In contrast, the WNN model captures the entire spectrum by simultaneously looking both forward and backward. Thus, the WNN layers do not contain padding and shrink at each layer by the dilation rate. In order to have the same output size as the input data, the circling feature of the frequency domain is used to double size of the input data.
            Fig. \ref{fig:DNN} illustrates this specific DNN architecture. 
            
            The presented design of the network requires an input spectrum consisting of $2^n-1$ points (Fig. \ref{fig:DNN}, gray circles in the input layer) that after the repetition (white circles in the input layer) form the input layer with $n-1$ hidden layers. Considering that size of the spectrum can be adjusted by the zero filling, the specific input size required by the DNN does not pose a limitation. Here, we used $2^{m-1}$ dilation rate for the $m$-th layer and 20 filters and $2\times1$ kernel size for each layer, with the ReLU activation functions between the layers.

            As described above, in the 2D case, a sliding window consisting of three spectra corresponding to three adjacent points along the direct dimension are processed together by WNN. These spectra represent three intertwined channels in DNN. As a result, DNN is trained to reconstruct three adjacent indirect NUS spectra simultaneously based on all three intertwined channels, like RGB channels are used in convolutional neural networks in image processing. Thus, the DNN, the input and output layers have three channels, where the adjacent spectra are simultaneously reconstructed. From these spectra only the middle is used in the final reconstruction.
            
        \subsubsection{Normalization}

            In order to train WNN for NUS reconstruction, $K$ pairs of input (NUS or J-coupled) $[s]_{N \times 3}^k$ and output (US or decoupled) $[S]_{N \times 3}^k$ are formed. Where, $k=1, 2, ..., K$ and each pair is composed of the three channels of the adjacent spectra. For each DNN in the WNN, both input and output are normalized using by maximum Euclidean norm value from the three channels.
            
        \subsubsection{Correction Step}
            WNN design (Fig. \ref{fig:WNN}) contains several DNNs (Fig. \ref{fig:DNN}) which optimal number depends on the NUS fraction. The initial spectrum $s_{NUS} $ containing strong aliasing artefacts feeds the first DNN, which reduces the artifacts and produces spectrum $S_{Pred}$ more similar to the uniform sampling spectrum. $S_{Pred}$ must be corrected before it is used as an input for next DNN, as described below. Similarly, Qu et al. \cite{Qu2019} previously stated that the FID at the location of sampled data points should be balanced between the acquired data points and the DNN-reconstructed data points. Here we used the following procedure to produce the corrected spectrum $S_{Cor}$: 
            
            \begin{equation}
               S_{Cor} = S_{Pred} - S_{Pred}* U + s_{NUS}
            \label{eq:cor}
            \end{equation}
            
            In the time domain, this correction is equivalent to restoring the experimental data in the FID, while keeping the predicted values for the not sampled points. As schematically shown in Fig. \ref{fig:WNN}, the correction improves the spectrum. Several consecutive DNN/correction steps are performed to reach good spectrum quality.
            
            As a note, to speed up the calculations, the convolution operation $S_{Pred}* U$ in Eq. \eqref{eq:cor} is performed in the time domain. Since the DNN deals only with the real part of the spectrum, we use the virtual echo representation \cite{Mayzel2014} to simplify transitions between the frequency and time domains.

    \subsection{The WNN Training}
        WNNs were trained on the NMRbox server \cite{NMRbox} (128 cores 2 TB memory), equipped with 4 NVIDIA A100 TENSOR CORE GPU graphics cards.
        We generated the network graphs using TensorFlow \cite{tensorflow2015-whitepaper} python package with the Keras frontend, and optimized DNNs within TensorFlow using the stochastic ADAM optimizer \cite{kingma2014adam} with the default parameters and 0.004 learning rate, Huber loss function, and the number of epochs and batch sizes of 500 and  128, respectively.
    
    \subsection{Protein NMR spectra}
        \subsubsection{Samples and Experiments}
        
        For the NMR experiments, we used the previously described protein samples: U \textemdash\ \ce{^{15}N}-\ce{^{13}C} \textemdash\ labeled Cu(I) azurin (14 kDa) \cite{korzhnev2003nmr}, U \textemdash\ \ce{^{15}N}-\ce{^{13}C} \textemdash\ labeled ubiquitin (8.6 kDa) \cite{brzovic2006ubch5}, and U \textemdash\ \ce{^{15}N}-\ce{^{13}C}-\ce{^{2}H} methyl ILV back-protonated MALT1 (44 kDa) \cite{unnerstaale2016backbone, malt12022met}.
        Fully sampled two-dimensional experiments used in this study are described in Table \ref{table:pra_spectra}.
        
        \begin{table*}[htbp]
            \centering
            \small
                \caption{Spectral parameters}
                \begin{tabular}{l|cccccc}
                    \hline
                    \hline
                    Protein & Size & Concentration & Spectrum & Spectral width & All points & NUS points\\
                     & (kDa) & (mM) & & (Hz) & & \\
                    \hline
                    Ubiquitin & 8.6 & 0.6 & \ce{^{1}H}-\ce{^{15}N} \textemdash\ HSQC & 3648.8 & 128 & 11 \\
                    Azurin & 14 & 1.0 & \ce{^{1}H}-\ce{^{15}N} \textemdash\ HSQC & 3648.8 & 128 & 21 \\
                    MALT1 & 44 & 0.5 & \ce{^{1}H}-\ce{^{15}N} \textemdash\ TROSY & 3283.9 & 128 & 31 \\
                    & & & \ce{^{1}H}-\ce{^{13}C} \textemdash\ HMQC & 4300.5 & 200 & - \\
                    & & & \ce{^{1}H}-\ce{^{13}C} \textemdash\ CT-HMQC & 4526.8 & 100 & - \\
                    \hline
                    \hline
                \end{tabular}
                \label{table:pra_spectra}
            \end{table*}
        
        \subsubsection{Processing with Compressed Sensing}
            We used python $nmrglue$ \cite{Nmrglue} and $NMRPipe$ \cite{nmrpipe} for reading and writing the NMR spectra and $mddnmr$ \cite{Jaravine_2006}. For the NUS processing with WNN and CS-IST \cite{Kazimierczuk2011}, the \ce{^{1}H}-\ce{^{15}N} correlation spectra were down-sampled to the number of NUS points specified in Table \ref{table:pra_spectra}. CS-IST and CS-virtual decoupling (CS-VD) \cite{Decoupling_K} calculations were performed using default $mddnmr$ parameters and the Virtual-Echo \cite{Mayzel2014} mode. 
        
        \subsubsection{NUS Spectra Quality Metrics}
            We used two metrics to assess the quality of the reconstructed NUS spectra with respect to the corresponding fully sampled spectra. These are the point-by-point root-mean-square deviation ($RMSD$) and the correlation coefficients ($R_S^2$). Before the comparison all spectra were normalized to their maximal peak intensity. To limit potential effects of the noise on the quality metrics, $RMSD$ and $R_S^2$ were calculated only for the spectral points with intensities above 1\% of highest peak intensity in either of the two compared spectra. Thus, both metrics are sensitive to the false-positive as well as false-negative spectral artefacts.
            
\section{Results and Discussion}
\label{sec:Results}

    \subsection{WNN 2D NUS Reconstruction Performance}
        Performance of the WNN's is demonstrated on several examples of \ce{^{1}H}-\ce{^{15}N} correlation spectra of different complexity for the three proteins Ubiquitin (8.6 kDa), Azurin (14 kDa), and MALT1 (44 kDa). We compared the quality of the reconstructed spectra obtained by WNNs trained using three different NUS/PSF schemes: fixed NUS schedule with defined PSF pattern (WNN-F), and two WNNs trained with varying NUS schedules, \textit{i.e.} Poisson-gap (WNN-P) and flat-random (WNN-R). The results were also compared with the spectra reconstructions by the Iterative Soft Thresholding algorithm (CS-IST) \cite{Kazimierczuk2011}, which is representative of the traditional NUS techniques implemented in $mddnmr$ software.
        
\begin{figure*}[htbp]
            \centering
            \includegraphics[width=0.85\textwidth]{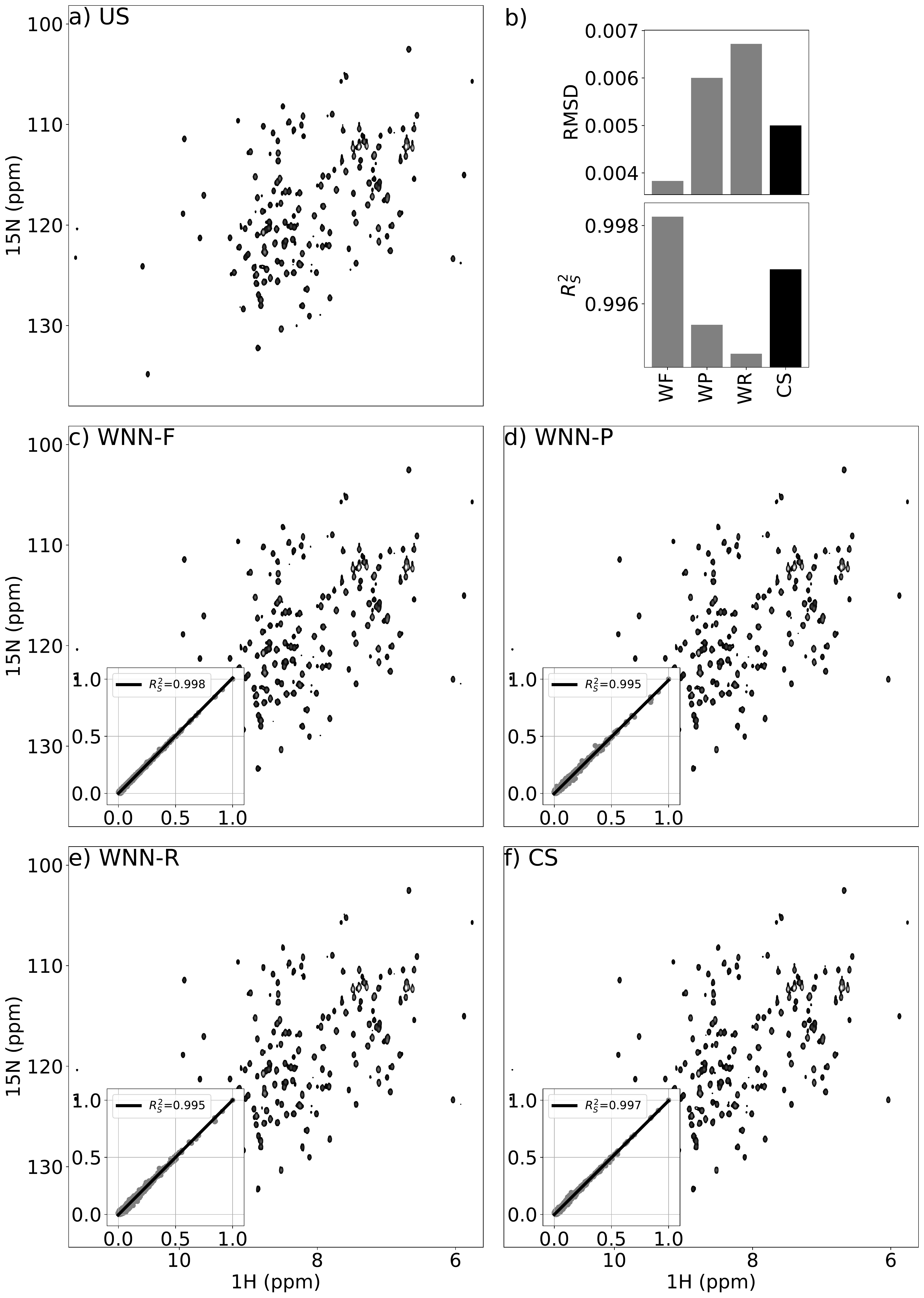}
            \caption{2D \ce{^{1}H}-\ce{^{15}N} \textemdash\ HSQC spectra of Azurin. (a) Uniformly Sampled (US) spectrum, (b) signal intensity RMSD and correlation coefficients ($R_S^2$) between the normalized US spectrum and spectra reconstructed using CS-IST (CS) and WNN's trained with: fixed Poisson-gap (WNN-F), unfixed Poisson-Gap (WNN-P), unfixed random sampling (WNN-R). See methods for details. (c-f) - Spectra reconstructed with WNN-F, WNN-P, WNN-R, and CS-IST, respectively. The insets show intensity correlations between the US and reconstructed spectra. }
            \label{fig:SPE_M}
        \end{figure*}

         Fig. \ref{fig:SPE_M} shows the reconstruction results for Azurin (results for Ubiquitin and MALT1 are shown in Fig. \ref{fig:SPE_S} and \ref{fig:SPE_L} in Appendix). It is clear that all four reconstruction methods, \textit{i.e.} WNN-F, WNN-P, WNN-R, and CS-IST, are capable of reproducing the spectrum with good and comparable quality using 21 NUS points (16.5\%) out of 128 points in the full reference spectrum. Comparison of the spectra quality metrics RMSD and $R_S^2$ (Fig. \ref{fig:SPE_M}b) reveals that among the three WNN's, the best result were obtained with WNN-F, which demonstrates the clear advantage of the fixed-schedule approach for WNN training. We therefore argue that while the WNN-F network is trained to recognize a specific PSF pattern produced by a defined schedule, the other two networks are trained on a multitude of different PSF realizations and thus cannot benefit from a "familiar" PSF pattern. 
         
\begin{figure}[htbp]
            \centering
            \begin{subfigure}[b]{0.45\textwidth}
                \centering
                \caption{}
                \includegraphics[width=\textwidth]{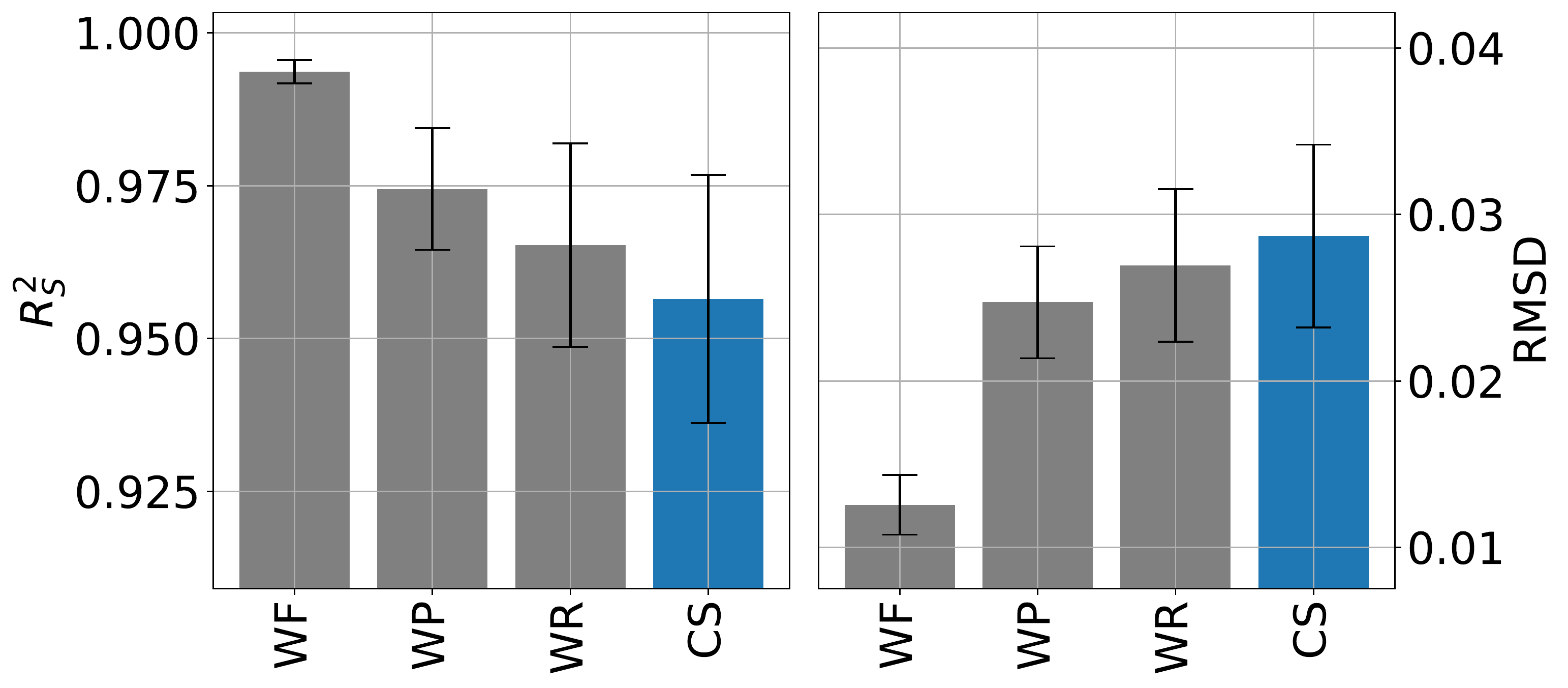}
                \label{fig:RR_D_S}
            \end{subfigure}
             \begin{subfigure}[b]{0.45\textwidth}
                \centering
                \caption{}
                \includegraphics[width=\textwidth]{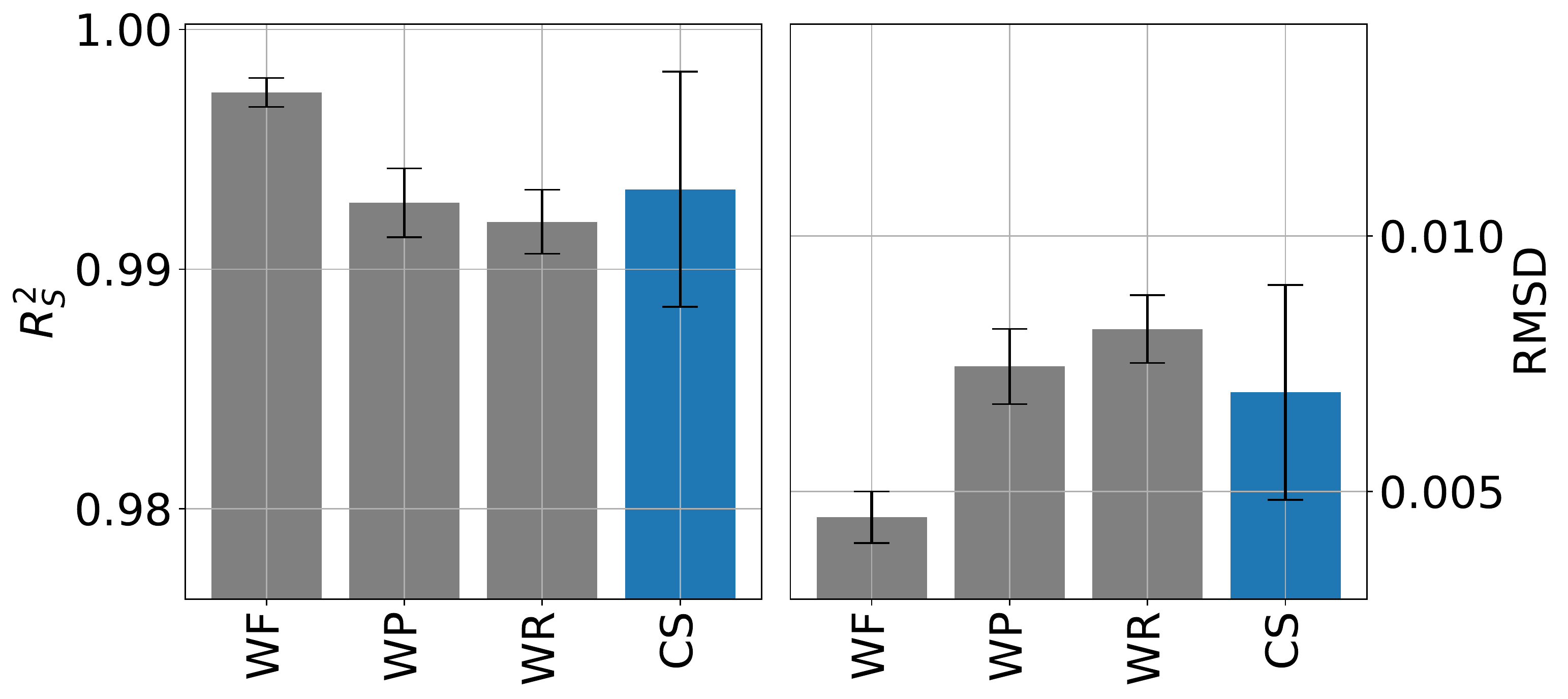}
                \label{fig:RR_D_M}
            \end{subfigure}
            \begin{subfigure}[b]{0.45\textwidth}
                \centering
                \caption{}
                \includegraphics[width=\textwidth]{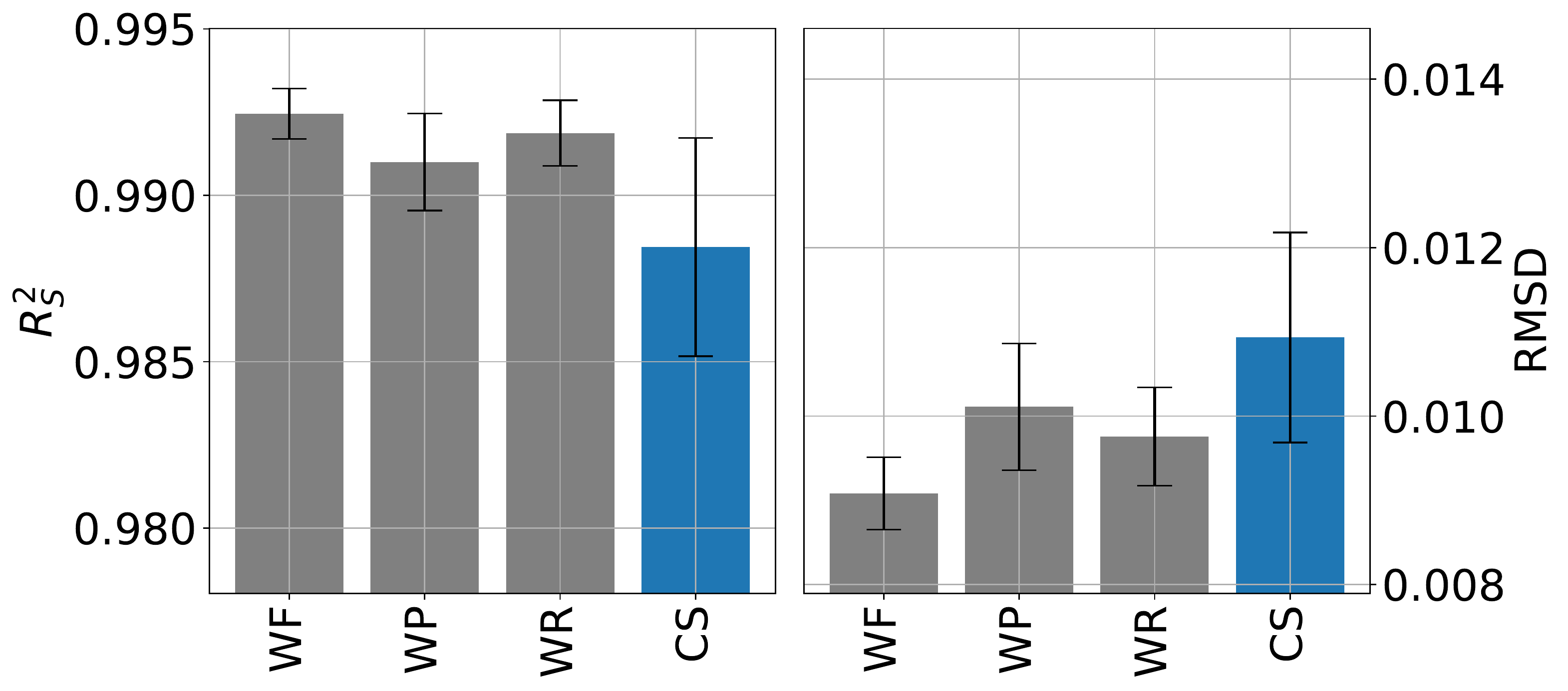}
                \label{fig:RR_D_L}
            \end{subfigure}
            \caption{Mean (bar graphs) and STD (error bar) of the signal intensity RMSD and correlation coefficients ($R_S^2$) between the normalized uniformly sampled and reconstructed NUS spectra using WNN trained with fixed (WNN-F), unfixed Poisson-Gap (WNN-P), and unfixed random (WNN-R) sampling and CS-IST (CS) (see Methods for details) (a) Ubiquitin spectra with a 11 out of 128 Poisson-gap sampling; (b) Azurin spectra with a 21 out of 128 Poisson-gap sampling; (c) MALT spectra with a 31 out of 128 Poisson-gap sampling.}
            \label{fig:RR_D}
        \end{figure}

        Figure \ref{fig:RR_D} shows RMSD and $R_S^2$ for the Ubiquitin, Azurin, and MALT1 proteins systems, which we obtained using 11, 21 , and 31 NUS points out of 128, respectively. In addition to the average, the error bars show spread of values over 11 different sampling schedules. With its higher $R_S^2$, lower RMSD, and smaller spreads of these values, the WNN-F displays significantly and consistently better results compared to the other WNN schemes and CS-IST. The main difference is the most pronounced for the low NUS fraction. We hypothesize that at the low NUS, the PSF aliasing artefacts are the strongest and constitute an easily detectable pattern for WNN-F. In contrast, at the higher NUS rate, intensities of the artefact peaks relative to the true/main peak are reduced, which reduces the value of the PSF pattern recognition and thus, diminishes the differences between the different WNN training schemes. The noticeable advantage of the WNN-P results over the WNN-R is consistent with previous observations \cite{huang2021exponential}. It is also noteworthy that although CS-IST produces reconstructions of similar average quality to WNN-P and WNN-R, it has a higher spread of the scores, which indicates a larger dependence of the results on the selected sampling schedule, especially at low NUS levels. 
        
    \subsection{$R^2$ and the number of WNNs}
\begin{figure}[htbp]
             \centering
             \includegraphics[width=0.45\textwidth]{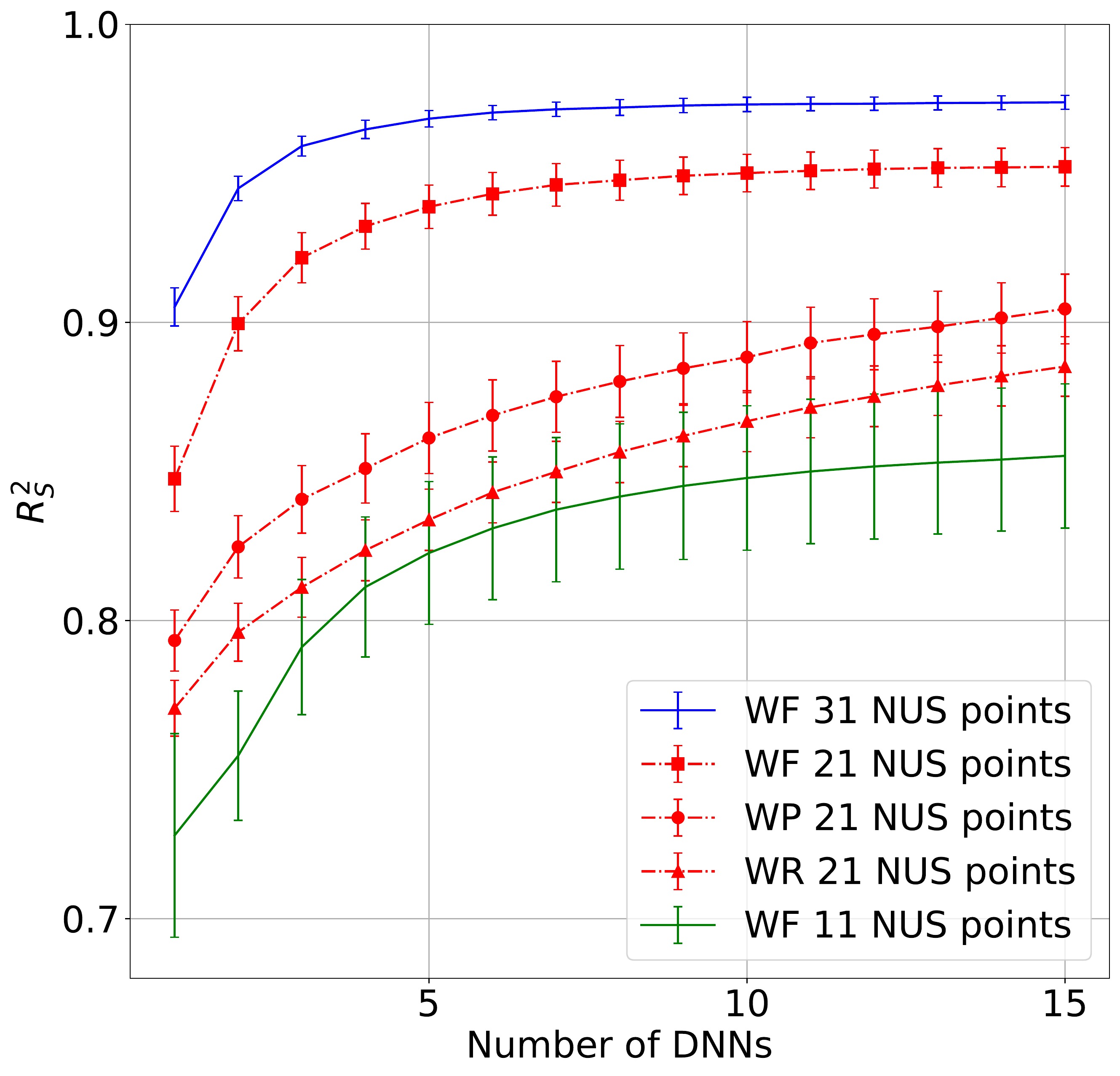}
        \caption{Mean and STD (error bar) versus number of DNNs in the WNN structure for the average signal intensity correlations $R^2$ between $1500 \times 11$ uniformly sampled synthetic spectra and corresponding reconstructed spectra using WNN-F (red squares), WNN-P (red circles) and WNN-R (red triangles) corresponding to a medium size protein spectrum with 21 points out of 128 NUS. (green) and (blue) - the results for WNN-F for small and large proteins with 11 and 31 NUS point out of 128, respectively.}
        \label{fig:D}
        \end{figure}

        As explained in the Method section and displayed in Figure \ref{fig:WDNN}, WNN consists of several nested DNNs which gradually improve the quality of the reconstructed spectrum. To reduce the amount of computations at the WNN training and spectra reconstruction stages, the number of DNNs should be small. Figure \ref{fig:D} presents statistics obtained on the 1500 synthetic test spectra corresponding to small, medium, and large proteins (see Section \ref{sec:Test data}). We measured average $R_S^2$ between 1500 uniform sampling synthetic spectra and the corresponding WNN-F reconstructed NUS spectra for small (green), medium (red) and large (blue) proteins with 11, 21, and 31 NUS point out of 128 respectively (Fig. \ref{fig:D}). Also, for medium proteins, in addition to WNN-F reconstructed spectra (red squares) we show $R_S^2$ results for WNN-P (red circles) and WNN-R (red triangles) reconstructions. Spread of the scores, which are shown as error bars, were obtained by repeating the calculations for 11 different NUS sampling schemes using the correspondingly trained WNN-F's. 
        The optimal number of DNNs, \textit{i.e.} when quality of the spectra does not improve with additional DNNs, depends primarily on the type of the sampling scheme as the WNN-F curves for 21 NUS points out of 128 show clear signs of leveling off at 6 DNN's, whereas WNN-P and WNN-R require at least 15 DNNs. Furthermore, comparison of the curves for the WNN-F at NUS levels 31/128, 21/128 and 11/128 show that fewer networks are needed for higher NUS fraction (Fig. \ref{fig:D}). 
        
    \subsection{WNN Robustness vs altered NUS schedule size}
\begin{figure}[htbp]
         \centering
             \begin{subfigure}[b]{0.45\textwidth}
                 \centering
                 \includegraphics[width=\textwidth]{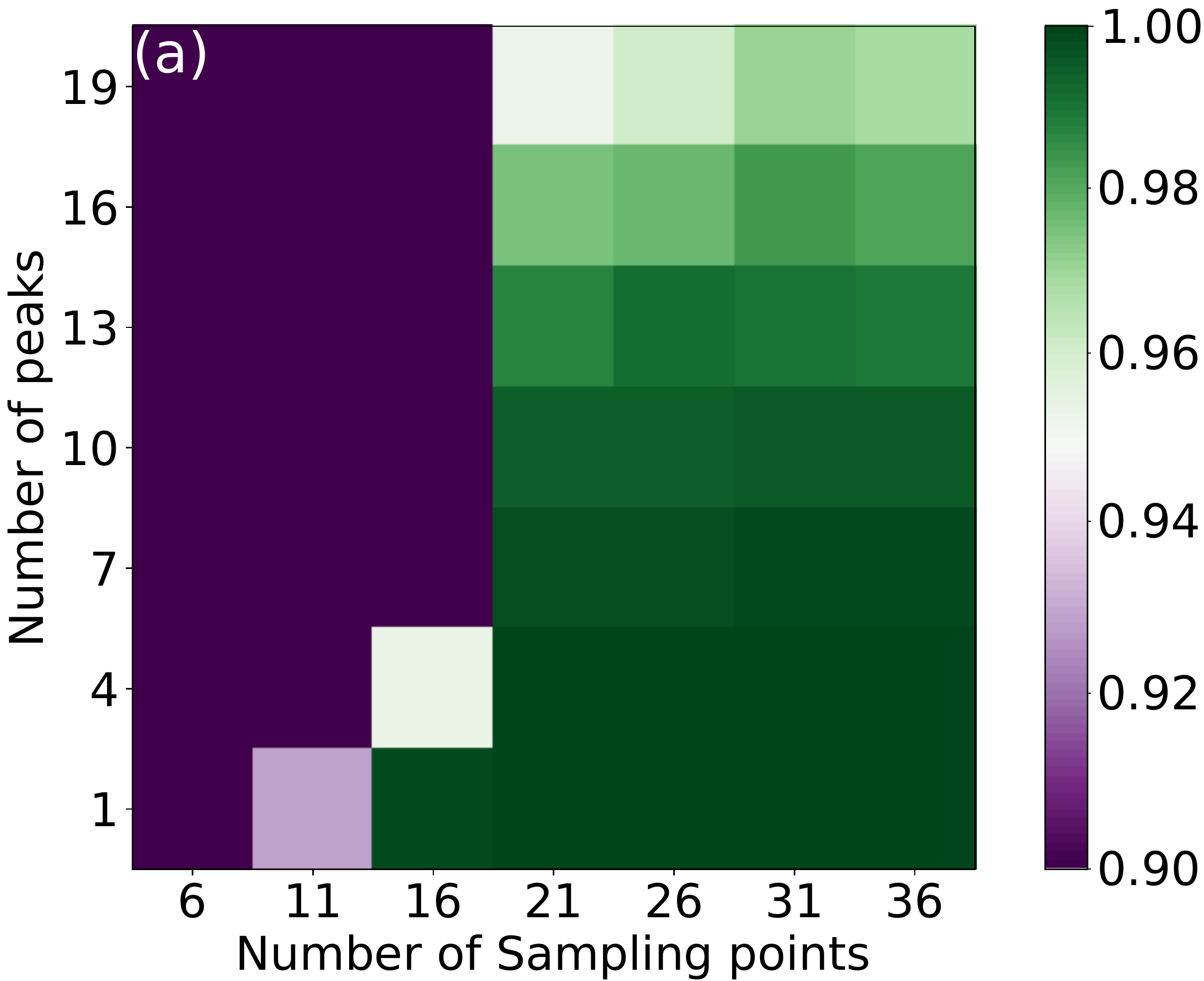}
             \end{subfigure}
             \begin{subfigure}[b]{0.45\textwidth}
                 \centering
                 \includegraphics[width=\textwidth]{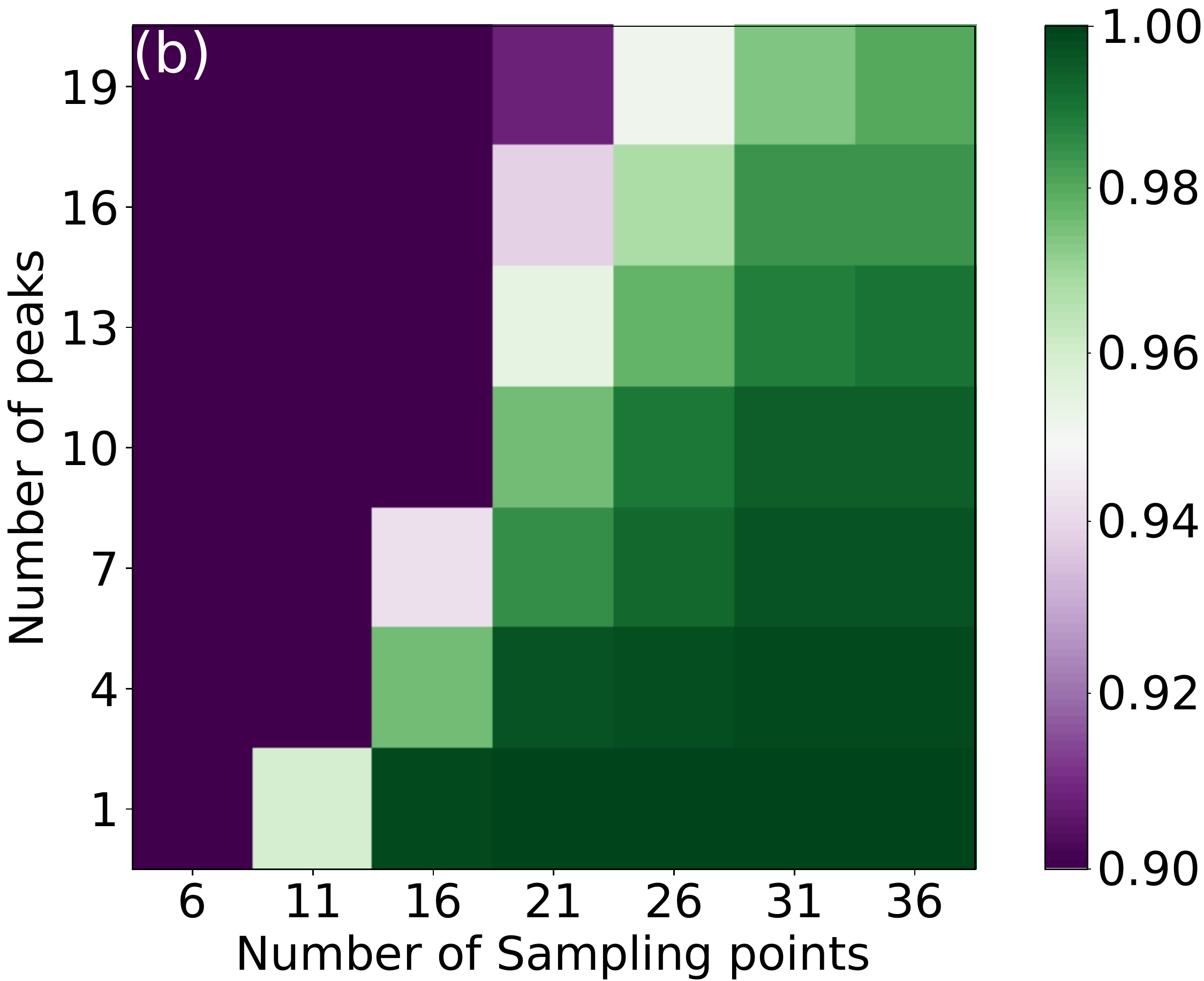}
             \end{subfigure}
             \begin{subfigure}[b]{0.45\textwidth}
                 \centering
                 \includegraphics[width=\textwidth]{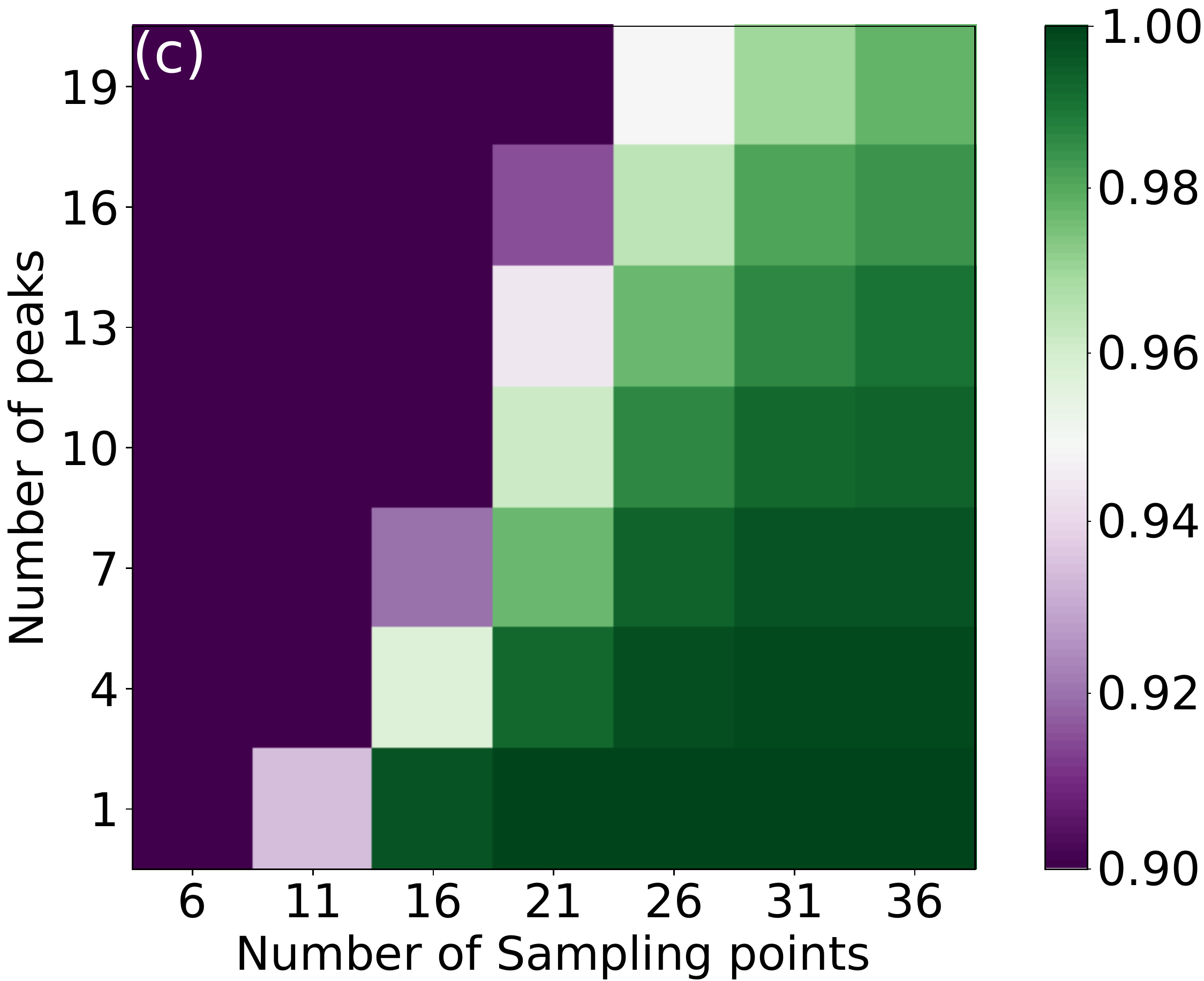}
             \end{subfigure}
        \caption{The average signal intensity correlations ($R_S^2$) between uniform sampling synthetic spectra and their corresponding reconstructed spectra using WNN-F (a), WNN-P (b) and, WNN-R (c) for a medium size protein (See Methods for details) when all three training data sets comprised 21 out of 128 NUS sampling rate.}
        \label{fig:R}
        \end{figure}

        All the WNN's were trained with defined NUS fractions, \textit{i.e.} 11, 21, 31 points out of 128. However, in practical experiments, it is often needed to adjust the number of points, for example to add more points to fit a spectrometer time allocation or to process an old experiment or an experiment that finished prematurely with fewer points. In these cases, the problem can be solved, although with relatively high computational costs, by training a new WNN with the actually needed NUS schedule. Figure \ref{fig:R} illustrates an alternative approach by showing performance of the WNN-F, WNN-P, and WNN-R beyond their training schedules size of 21 points. An altered schedule is produced by either truncating the NUS table or by augmenting its size with additional random points. Thereafter, corresponding test spectra were simulated (see Section \ref{sec:Test data}). All WNN's types produce better $R_S^2$ scores with more points than in the training data. WNN-F shows the best result for 21 points and 26, when it can reliably reproduce spectra with up to 19 peaks. With 31 and 36 points, $R_S^2$ scores are even better, while WNN-F and WNN-P are similar. For all WNN's, the scores drop sharp when less than 21 NUS points are used for the spectra reconstruction. This indicates that 21 point represents the lower border line for successful reconstruction. The comparatively lower score for WNN-F when using less than 21 NUS point may be explained by fast degradation of the specific PSF pattern when points are removed from the schedule. These results demonstrate that all WNNs can be used with more NUS points than for their training, but reconstructions with less points should be performed with caution. 
        
        The simulations confirm the result presented for the real spectra in Figures \ref{fig:SPE_M}, \ref{fig:RR_D},  \ref{fig:SPE_S}, and \ref{fig:SPE_L} that show significantly higher quality spectra reconstructions by WNN-F compared to the other processing schemes. For the WNNs, this is easily rationalized since a network obviously benefits from training on the exactly the same PSF as it is used in the presented experiment. In the broader context, CS-IST, WNN-P and WNN-R, which do not take advantage of prior knowledge about the PSF, displays similar performances. Although sometimes use of a fixed NUS schedule may be considered as a luck of flexibility and thus a disadvantage \cite{Karunanithy2021,Hansen2019}, based on our results, we hypothesized that WNN-F ability to outperform the other methods is due to it making use of PSF as new, so far untapped, prior for the successful spectra reconstruction. 
        
    \subsection{Virtual Decoupling by WNN}
    \label{sec:VD}
    
\begin{figure*}[htbp]
        \centering
        \includegraphics[width=0.95\textwidth]{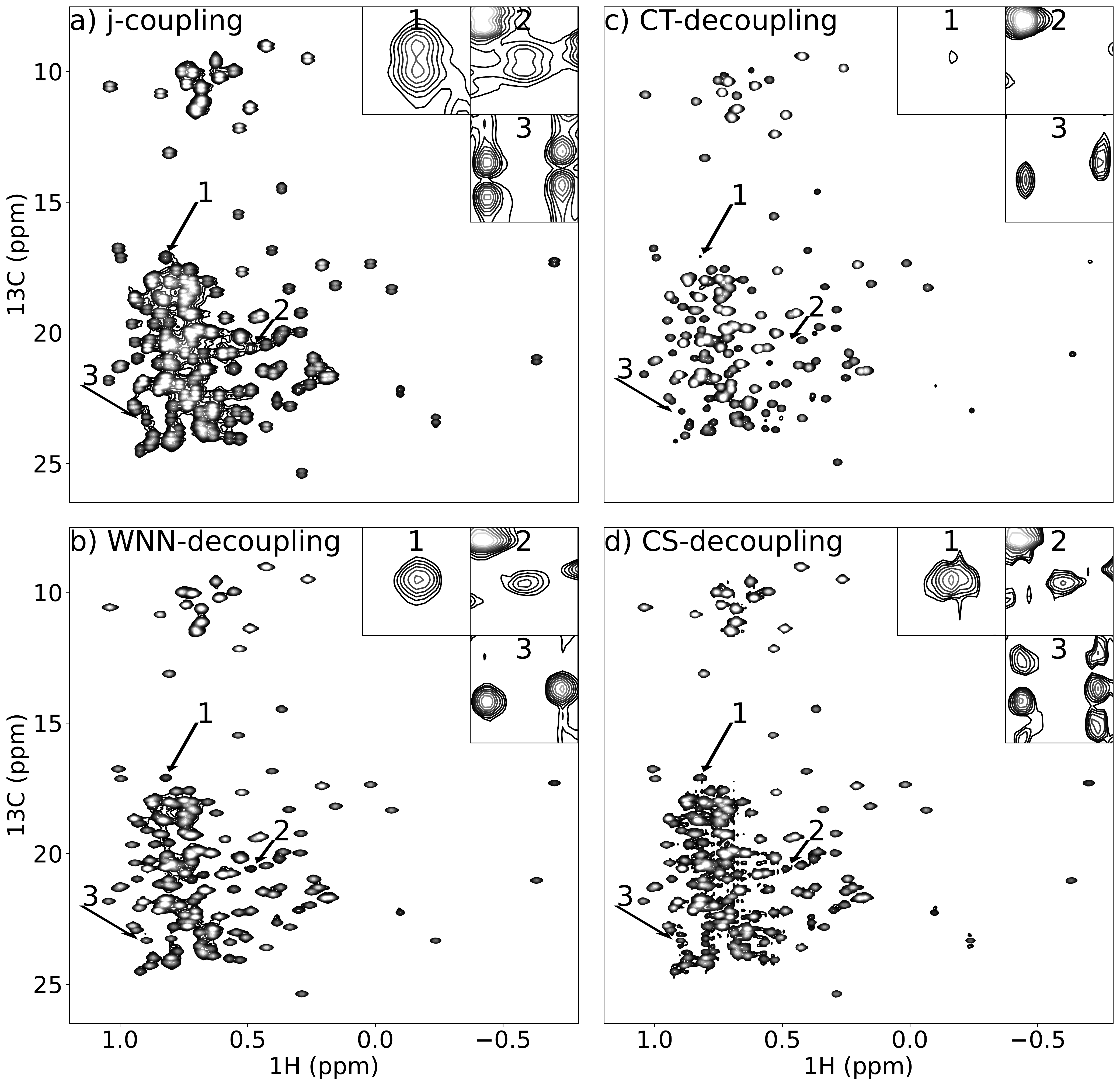}
        \caption{High resolution methyl 2D \ce{^{1}H}-\ce{^{13}C} \textemdash\ HMQC spectra of protein MALT1. Each spectrum is normalized and plotted at the same contour level with only positive contours shown. (a) Spectrum without decoupling processed with Fourier transform. (b) Spectrum decoupled using WNN-F. (c) Constant time \ce{^{1}H}-\ce{^{13}C} \textemdash\ HMQC spectrum. (d) Spectrum decoupled by deconvolution using the CS algorithm.}
        \label{fig:SPE_VD}
    \end{figure*}

    Similar to PSF, the WNN architecture can be trained to recognize other patterns in the spectra. The obvious examples are peak multiplets caused by scalar coupling. In this work, we demonstrate virtual homo decoupling in methyl 2D \ce{^{1}H}-\ce{^{13}C} \textemdash\ HMQC spectrum of 44 kDa protein MALT1 (Fig. \ref{fig:SPE_VD}). Without the decoupling (Fig. \ref{fig:SPE_VD}a), the \ce{^{13}C} resonances are split due to approximately 35 Hz J-coupling with the adjacent methyl \ce{^{13}C} atom. This drastically reduces the spectral resolution. In the experiment, the peak splitting can be suppressed by using constant-time (CT) evolution, albeit with the price of significant loss of sensitivity, especially for large molecular systems. Effects of the CT decoupling are illustrated in Fig. \ref{fig:SPE_VD}c. Example of peaks enlarged in inset 3 are narrowed at the expense of significant attenuation and even loss of the peaks shown in insets 1 and 2. 
    
    The virtual decoupling (VD) may solve the sensitivity problem (Fig. \ref{fig:SPE_VD}b,d). Furthermore, VD allows for higher flexibility in choosing the acquisition time than the CT and, thus may offer higher practical spectral resolution. However, VD using traditional algorithmic techniques \cite{bothner1987useful,delsuc1988application,serber2000new,kerfah2015ch3,Decoupling_K} as exemplified by  CS-IST in Fig. \ref{fig:SPE_VD}d have a caveat in necessity to provide the algorithm with nearly exact value of the coupling constant. Two peaks shown in insets 3 in Figure \ref{fig:SPE_VD} have J-coupling of 41 Hz, which is larger than the value (35 Hz) used for the whole spectrum. As a result, peaks in inset 3 in Figure \ref{fig:SPE_VD}d are corrupted. Our WNN was trained for a range of J-coupling values and thus demonstrates excellent virtual decoupling (Fig. \ref{fig:SPE_VD}b) with both high sensitivity and tolerance to the dispersion of the J-coupling values. Although, excellent pattern recognition ability of DNN's have been utilized for the VD \cite{Hansen2019,karunanithy2021virtual}, our results point to the fundamental similarity of the problems of the VD and spectral reconstruction from NUS data. In both cases, known patterns of the spectral features, \textit{i.e.} PSF and peak multiplets, can be used as a valuable prior knowledge to successfully reconstruct the spectrum. 
    
\section{Conclusion}
\label{sec:Conclusion}
    We present a new DNN-based architecture WNN, which is specifically designed to grasp patterns over the entire NMR spectrum. If trained at a fixed NUS schedule, the WNN benefits from pattern recognition of the corresponding PSF pattern produced by each peak, which allows highest quality and robust reconstruction of the NUS spectra. As another example of the pattern recognition by WNN, we demonstrate virtual decoupling in 2D methyl \ce{^{1}H}-\ce{^{13}C} \textemdash\ HMQC spectrum of 44 kDa protein MALT1. As far as we know, WNN is the first tool that specifically uses PSF as a prior knowledge in NUS spectra reconstruction. WNN demonstrates that the pattern-oriented signal processing schemes may be very efficient and surpass the existing algorithmic methods developed for the reconstruction of NUS spectra, which do not fully exploit the knowledge about the NUS schedule. 

\section*{Acknowledgement}
    This study used NMRbox: National Center for Biomolecular NMR Data Processing and Analysis, a Biomedical Technology Research Resource (BTRR), which is supported by NIH grant P41GM111135 (NIGMS). The work was supported by the Swedish Foundation for Strategic Research grant ITM17-0218 to T.A and P.A., grant RSF 19-74-30014 to D.M.L., Swedish Cancer Society grant 21 1605 Pj01H to A.A., and the Swedish Research Council grants 2021-05061 to A.A. and 2019-03561 to V.O. 

\twocolumn

\bibliographystyle{ieeetr} 
\bibliography{references}

\newpage
\onecolumn

\begin{appendices}
\section{Appendix}
\label{appendix}
\setcounter{figure}{0}
\counterwithin{figure}{section}

    \begin{figure}[htbp]
        \centering
        \includegraphics[width=0.65\textwidth]{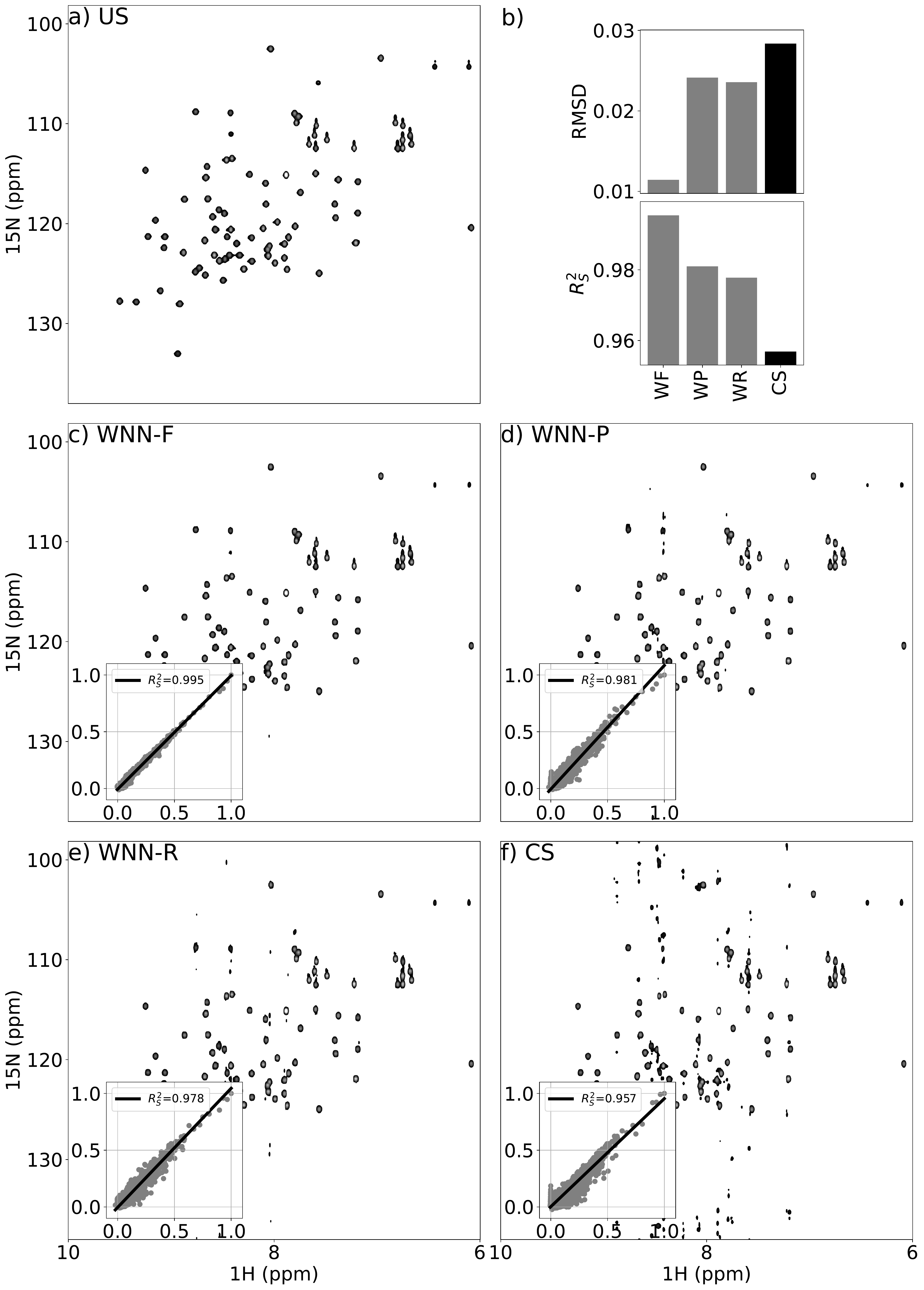}
        \caption{2D \ce{^{1}H}-\ce{^{15}N} \textemdash\ HSQC spectra of Ubiquitin. (a) Uniformly Sampled (US) spectrum, (b) signal intensity RMSD and  correlation coefficients ($R_S^2$) between the normalized US spectrum and spectra reconstructed using CS-IST (CS) and WNN's trained with: fixed Poisson-gap (WNN-F), unfixed Poisson-Gap (WNN-P), unfixed random sampling (WNN-R). See methods for details. (c-f) - Spectra reconstructed with WNN-F, WNN-P, WNN-R, and CS-IST, respectively. The insets show intensity correlations between the US and reconstructed spectra.}
        \label{fig:SPE_S}
    \end{figure}
    
    \begin{figure*}[htbp]
        \centering
        \includegraphics[width=0.65\textwidth]{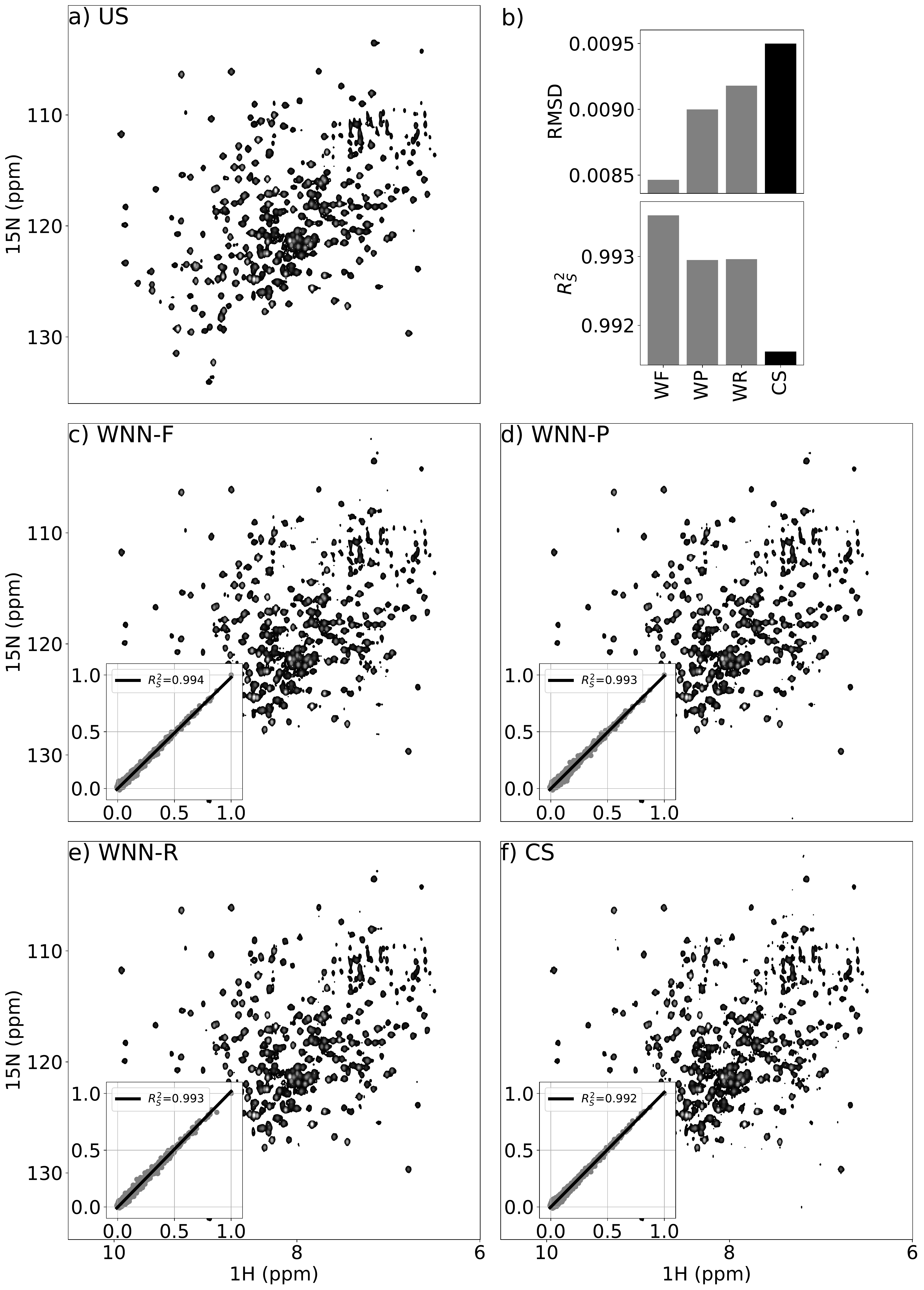}
        \caption{2D \ce{^{1}H}-\ce{^{15}N} \textemdash\ TROSY spectra of MALT1. (a) Uniformly Sampled (US) spectrum, (b) signal intensity RMSD and  correlation coefficients ($R_S^2$) between the normalized US spectrum and spectra reconstructed using CS-IST (CS) and WNN's trained with: fixed Poisson-gap (WNN-F), unfixed Poisson-Gap (WNN-P), unfixed random sampling (WNN-R). See methods for details. (c-f) - Spectra reconstructed with WNN-F, WNN-P, WNN-R, and CS-IST, respectively. The insets show intensity correlations between the US and reconstructed spectra.}
        \label{fig:SPE_L}
    \end{figure*}
\end{appendices}
\end{document}